# Twelve-Dimensional Supersymmetric Gauge Theory as the Large $N$ Limit [1]

Hitoshi NISHINO

*Department of Physics*
*University of Maryland at College Park*
*College Park, MD 20742-4111, USA*
E-Mail: nishino@nscpmail.physics.umd.edu

## Abstract

Starting with the ordinary ten-dimensional supersymmetric Yang-Mills theory for the gauge group $U(N)$, we obtain a twelve-dimensional supersymmetric gauge theory as the large $N$ limit. The two symplectic canonical coordinates parametrizing the unitary $N \times N$ matrices for $U(N)$ are identified with the extra coordinates in twelve dimensions in the $N \to \infty$ limit. Applying further a strong/weak duality, we get the 'decompactified' twelve-dimensional theory. The resulting twelve-dimensional theory has peculiar gauge symmetry which is compatible also with supersymmetry. We also establish a corresponding new superspace formulation with the extra coordinates. By performing a dimensional reduction from twelve dimensions directly into three dimensions, we see that the Poisson bracket terms which are needed for identification with supermembrane action arises naturally. This result indicates an universal duality mechanism that the 't Hooft limit of an arbitrary supersymmetric theory promotes the original supersymmetric theory in $(D-1, 1)$ dimensions into a theory in $(D, 2)$ dimensions with an additional pair of space-time coordinates. This also indicates interesting dualities between supermembrane theory, type IIA superstring with $D0$-branes, and the recently-discovered twelve-dimensional supersymmetric theories.

---

[1]This work is supported in part by NSF grant # PHY-93-41926.

# 1. Introduction

There has been accumulating evidence that $U(N)$ matrix theory [1] in the large $N$ limit [2] corresponds to the strongly coupled type IIA superstring [3], and therefore to the M-theory [4][5]. For example, we can explicitly compute the supergraviton effective potential that agrees with the that of eleven-dimensional (11D)[^2] supergravity theory [6].

In the $N \to \infty$ limit of an $U(N)$ Yang-Mills (YM) theory, the commutators for $N \times N$ matrices $Z$ and $W$ are replaced by the Poisson bracket $\{Z, W\}_P \equiv (\partial_p Z)(\partial_q W) - (\partial_q Z)(\partial_p W)$ where $p$ and $q$ are commuting variables in the large $N$ limit, which were originally non-commuting variables, satisfying $[q, p] = 2\pi i/N$. It has been also conjectured [7] that the large $N$ limit of $U(N)$ YM theories with 16 supercharges are related to certain supergravity solutions. These recent developments indicate an universal mechanism relating a supersymmetric theory in $(D-1, 1)$ dimensions with $(D, 2)$ dimensions involving two extra coordinates $p$ and $q$, replacing all the non-Abelian commutators by the Poisson brackets in $(D, 2)$ dimensions. Our formulation is in a sense similar to the formulation in [8] for supermembrane action, in the sense that the Poisson bracket terms are identified with the large $N$ limit of $U(N)$ supersymmetric YM theory.

In this paper, we study if the mechanism as above works even in the case of 10D supersymmetric $U(N)$ YM theory in the $N \to \infty$ limit, with all the non-Abelian group commutators replaced by the Poisson brackets with respect to the extra coordinates in 12D. Interestingly, we will find that the resulting 12D theory has a peculiar 'gauge' symmetry, which is also consistent with 12D supersymmetry, similar to a previous formulation of 12D supersymmetric YM theory [9]. Our formulation here is similar to that in [10], in which $Sp(2, \mathbb{R})$ symmetry for the position/momentum is treated as a local symmetry embedded in $SO(d, 2)$. However, our formulation is also different in the sense that we do not deal with point particle action with bi-local fields, and $Sp(2, \mathbb{R})$ is not gauged, either. After establishing the $N \to \infty$ action, we perform a dimensional reduction of our 12D theory into 3D, in order to compare the result with the supermembrane action. As desired, we see that the resulting action agrees with the action obtained by $N \to \infty$ limit of the $D0$-brane action in 1D [1], and therefore, it coincides with the supermembrane action in 3D.

# 2. Canonical Variables $p$ and $q$ and Large $N$ Limit

We first review the parametrization of $U(N)$ $N \times N$ matrices in terms of canonical variables $p$ and $q$ [11][12][1], and its effect on the commutators in the large $N$ limit [2].

[^2]: We use 11D or $D = 11$ for eleven dimensions, when the signatures are not crucial. To specify the signatures, we use $(s, t)$ for $s$ positive space and $t$ negative time signatures.

Any $N \times N$ complex matrix $Z$ can be expanded in terms of two unitary $N \times N$ matrices $U$ and $V$, satisfying

$$U^N = I \ , \quad V^N = I \ , \quad UV = e^{2\pi i/N} VU \ , \tag{2.1a}$$

$$U = e^{ip} \ , \quad V = e^{iq} \ , \quad [q,p] = \tfrac{2\pi i}{N} \ , \tag{2.1b}$$

with so-called canonical variables $p$ and $q$, as

$$Z = \sum_{m,n=0}^{N-1} z_{mn} U^m V^n = \sum_{m,n=0}^{N-1} z_{mn} e^{imp} e^{inq} \ , \quad z_{mn} = \tfrac{1}{N} \operatorname{tr}(U^{-m} Z V^{-n}) \ . \tag{2.2}$$

Eq. (2.1) implies that the eigenvalues of $p$ and $q$ can be chosen to be $-\pi,\ -\pi(N-1)/N,\ -\pi(N-2)/N,\ \cdots,\ -\pi/N,\ 0,\ \pi/N,\ 2\pi/N,\ \cdots,\ (N-1)\pi/N$.[3]

In the large $N$ limit [2], $p$ and $q$ become mutually commuting $c$-numbers, as (2.1b) shows. Moreover, the eigenvalues of $p$ and $q$ become continuous taking all the real values in $-\pi \leq p < \pi,\ -\pi \leq q < \pi$, behaving like a pair of coordinates for a phase space [11][12][1]. In such a limit, the $Z$ in (2.2) becomes just an ordinary Fourier expansion in terms of $p$ and $q$, which we call $z(p,q)$:

$$z(p,q) \equiv \sum_{m,n=0}^{\infty} z_{mn} e^{imp} e^{inq} \ , \tag{2.3a}$$

$$z_{mn} = \int_{-\pi}^{\pi} \frac{dp}{2\pi} \int_{-\pi}^{\pi} \frac{dq}{2\pi} \ z(p,q) e^{-imp-inq} \ . \tag{2.3b}$$

Accordingly, we have[4]

$$\operatorname{tr} Z \to N \int_{-\pi}^{\pi} \frac{dp}{2\pi} \int_{-\pi}^{\pi} \frac{dq}{2\pi} \ z(p,q) \ , \tag{2.4a}$$

$$[Z,W] \to \tfrac{2\pi i}{N} (Z_{,q} W_{,p} - Z_{,p} W_{,q}) \ . \tag{2.4b}$$

Eq. (2.4b) implies that an $U(N)$ commutator can become a Poisson bracket in the large $N$ limit. Note that due to the symplectic feature of these two coordinates, it is natural to have the indefinite signature $(+,-)$ in the $(p,q)$ space, and it is convenient to use the coordinates $(x^+, x^-) \equiv \left((p+q)/\sqrt{2},\ (p-q)/\sqrt{2}\right)$.

In principle, we can apply this aspect of the large $N$ limit for $U(N)$ to any YM theory in any dimensions. For example, after appropriate rescalings by the powers of $N$, the $U(N)$ YM field strength

$$F_{\mu\nu} \equiv A_{\nu,\mu} - A_{\mu,\nu} + ig[A_\mu, A_\nu] \ , \tag{2.5}$$

---

[3]We use this convention instead of $0,\ \pi/N,\ 2\pi/N,\ \cdots, 2(N-1)\pi/N$ in [1][5] for a later purpose of decompactification.

[4]In this paper we use the symbol like $_{,p}$ to denote the derivative $\partial/\partial p$. We avoid the usage of $\partial_\mu$ for $\partial/\partial x^\mu$, because $A_{\mu,+} A_{\nu,-}$ *etc.*, are more compact than $(\partial_+ A_\mu)(\partial_- A_\nu)$, in equations like (2.6).

in $(D-1,1)$ dimensions with the coordinates $(x^0, x^1, \cdots, x^{D-1})$ can be promoted into the field strength

$$F_{\mu\nu} \equiv A_{\nu,\mu} - A_{\mu,\nu} + g\left(A_{\mu,+}A_{\nu,-} - A_{\mu,-}A_{\nu,+}\right) \ , \tag{2.6}$$

in the $(D,2)$ dimensions $(x^0, x^1, \cdots, x^{D-1}, x^+, x^-)$. The constant $g$ in (2.5) is the usual YM coupling constant. The last term in (2.6) is nothing else than the Poisson bracket replacing the commutator when $N$ is finite. Even though (2.6) seems rather unusual with the last term, we will see shortly how this makes sense as a field strength, transforming properly under our gauge transformation. The metric of the resulting $(D,2)$ dimensions is $(\eta_{\mu\nu}) = \text{diag.}(-,+,+,\cdots,+,-)$ with an additional pair of space-time coordinates, equivalent to the symplectic variables $p$ and $q$.

The geometrical meaning of this process is clear, from the viewpoint that the two variables $p$ and $q$ can be interpreted as coordinates of particles in quantum mechanics [1]. By adding two additional coordinates to the usual base manifold, the total space-time dimensions become now two dimensions higher than the original one.[5] As in (2.3), the range for the new variables $p$, $q$ is to be $[-\pi, \pi)$. This restriction of coordinates implies nothing else than 'compactification' on $S^1 \otimes S^1$ of the extra dimensions. In order to get 'decompactified' system with the extra coordinates free of such a restriction, we need to adopt additional limiting procedure based on strong/weak duality. This can be done as follows. Consider the rescaling of the extra coordinates $x^{D+1}$ and $x^{D+2}$ by

$$y^{D+1} = Rx^{D+1} \ , \quad y^{D+2} = Rx^{D+2} \quad (-\pi R \leq y^{D+1} < \pi R, \ -\pi R \leq y^{D+2} < \pi R) \ , \tag{2.7}$$

and take the limit

$$R \to \infty \ , \quad g \to 0 \quad \text{with} \quad \widetilde{g} \equiv R^2 g \quad \text{fixed.} \tag{2.8}$$

Now the field strength (2.6) stays formally the same, except that the derivatives $_{,\pm}$ are now with respect to the new coordinates $y^\pm$ with the ranges $-\infty < y^{D+1} < \infty$, $-\infty < y^{D+2} < \infty$, and that the coupling constant $g$ is now replaced by $\widetilde{g}$. In other words, by taking this particular limit, we can realize the 'decompactification' from $(\text{Minkowski})_{10} \otimes S^1 \otimes S^1$ into $(\text{Minkowski})_{12}$. In the next section, we apply this prescription to the usual 10D supersymmetric YM theory for the gauge group $U(N)$, to get the decompactified 12D theory. Accordingly, all the field strengths used from now on are understood to be in terms of the new decompactified coordinates $-\infty < y^{D+1} < \infty$, $-\infty < y^{D+2} < \infty$, and the rescaled coupling constant $\widetilde{g}$, even though we use their original symbols like $x^{D+1}$, $x^{D+2}$ and $g$, in order to simplify the notation.

---
[5]The geometrical significance and consistency with supersymmetry will be more elucidated, when we reformulate in superspace in a later section.

## 3. Supersymmetric Gauge Theory in 12D with Peculiar Gauge Symmetry

We first summarize our result on our 12D supersymmetric gauge theory after the large $N$ limit and our decompactification limit, which is very similar to [9], but with a peculiar gauge symmetry arising from the $N \to \infty$ limit. As was mentioned, we need two time directions for promoting 10D supersymmetric YM theory to 12D, due to the two symplectic variables.

Our notation is the same as the component formulation in [13], *i.e.*, our metric is $(\eta_{\mu\nu}) =$ diag. $(-, +, \cdots, +, +, -)$, where we use the indices $\mu, \nu, \cdots = $ 0, 1, $\cdots$, 9, 11, 12 for the 12D coordinates, Accordingly, our Clifford algebra is $\{\gamma_\mu, \gamma_\nu\} = +2\eta_{\mu\nu}$, with $\epsilon^{012\cdots 9\,11\,12} = +1$, and $\gamma_{13} \equiv \gamma_0 \gamma_1 \cdots \gamma_9 \gamma_{11} \gamma_{12}$. We use null-vectors [9] defined by

$$(n^\mu) = (0, 0, \cdots, 0, +\tfrac{1}{\sqrt{2}}, -\tfrac{1}{\sqrt{2}}) \ , \qquad (n_\mu) = (0, 0, \cdots, 0, +\tfrac{1}{\sqrt{2}}, +\tfrac{1}{\sqrt{2}}) \ ,$$
$$(m^\mu) = (0, 0, \cdots, 0, +\tfrac{1}{\sqrt{2}}, +\tfrac{1}{\sqrt{2}}) \ , \qquad (m_\mu) = (0, 0, \cdots, 0, +\tfrac{1}{\sqrt{2}}, -\tfrac{1}{\sqrt{2}}) \ . \quad (3.1)$$

We also use $\pm$-indices [14][15], for the two extra dimensions in 12D: $V_\pm \equiv 2^{-1/2}(V_{(11)} \pm V_{(12)})$. It then follows that $n_+ = m^+ = +1$, $n_- = m^- = 0$, and therefore

$$n^\mu n_\mu = m^\mu m_\mu = 0 \ , \qquad m^\mu n_\mu = m^+ n_+ = m_- n^- = +1 \ . \quad (3.2)$$

The $P_\uparrow$, $P_\downarrow$ are our projection operators for the space of extra dimensions, satisfying the ortho-normality conditions [14]:

$$P_\uparrow \equiv \tfrac{1}{2} \slashed{n} \slashed{m} = \tfrac{1}{2} \gamma^+ \gamma^- \ , \qquad P_\downarrow \equiv \tfrac{1}{2} \slashed{m} \slashed{n} = \tfrac{1}{2} \gamma^- \gamma^+ \ , \qquad P_\uparrow P_\downarrow = P_\downarrow P_\uparrow = 0 \ ,$$
$$P_\uparrow P_\uparrow = +P_\uparrow \ , \qquad P_\downarrow P_\downarrow = +P_\downarrow \ , \qquad P_\uparrow + P_\downarrow = +I \ , \quad (3.3)$$

with $\slashed{m} \equiv m^\mu \gamma_\mu$ and $\slashed{n} \equiv n^\mu \gamma_\mu$. The symmetry of the $\gamma$-matrices are such as

$$(\slashed{n})_{\alpha\dot{\beta}} = -(\slashed{n})_{\dot{\beta}\alpha} \ , \qquad (\slashed{m})_{\alpha\dot{\beta}} = -(\slashed{m})_{\dot{\beta}\alpha} \ , \qquad (P_\uparrow)_{\alpha\beta} = -(P_\downarrow)_{\beta\alpha} \ , \quad (3.4)$$

where undotted spinorial indices $\alpha, \beta, \cdots = $ 1, 2 are for the negative chiral components, while $\dot{\alpha}, \dot{\beta}, \cdots = \dot{1}, \dot{2}$ are for positive chirality.[6]

The field content for our 12D theory is $(A_\mu, \lambda)$, where $\lambda$ is a Majorana-Weyl spinor satisfying $\gamma_{13} \lambda = -\lambda$. Our total action in 12D, obtained by the prescription in the last section, is now

$$I \equiv \int d^{12}x \left[ -\tfrac{1}{4}(F_{\mu\nu})^2 + \tfrac{1}{2} F_{\mu\nu} F^{\mu\rho} n^\nu m_\rho + \tfrac{1}{2}(F_{\mu\nu} m^\mu n^\nu)^2 + (\overline{\lambda} P_\uparrow \gamma^\mu \slashed{n} D_\mu \lambda) \right] \ , \quad (3.5)$$

---
[6]We follow refs. [9][14] for the *dottedness* of indices, which is opposite to the usual convention.

where our field strength and 'covariant' derivative are defined by

$$F_{\mu\nu} \equiv A_{\nu,\mu} - A_{\mu,\nu} + g(A_{\mu,+}A_{\nu,-} - A_{\mu,-}A_{\nu,+}) ,$$
$$D_\mu \lambda \equiv \lambda_{,\mu} + g(A_{\mu,+}\lambda_{,-} - A_{\mu,-}\lambda_{,+}) . \quad (3.6)$$

Since these quantities are understood as the large $N$ limit, there is no 'hidden' index like adjoint indices in 10D, and there is no need to take the trace in (3.5). As has been mentioned at the end of section 2, the ranges of the extra coordinates are $-\infty < x^{11} < \infty$, $-\infty < x^{12} < \infty$, after our decompactification limit. Therefore there is no difference about the range of coordinates between the 10D ones and the extra ones.

The $\pm$-derivative terms in (3.6) are identified with the Poisson brackets with respect to the extra coordinates $\pm$ as a reminiscent of the non-Abelian commutators for the adjoint representation of $U(N)$ in 10D.

Our supersymmetry transformation rule is similar to [16] with a slight difference:

$$\delta_Q A_\mu = (\bar{\epsilon} P_\uparrow \gamma_\mu \displaystyle{\not}n \lambda) = (\bar{\epsilon} \gamma_i \displaystyle{\not}n \lambda) \delta_\mu{}^i , \quad (3.7a)$$
$$\delta_Q \lambda = +\tfrac{1}{4} P_\downarrow \gamma^\mu P_\downarrow \gamma^\nu P_\downarrow \epsilon F_{\mu\nu} = \tfrac{1}{4} P_\downarrow \gamma^{ij} \epsilon F_{ij} , \quad (3.7b)$$

where the indices $i, j, \cdots = 0, 1, \cdots, 9$ are for the purely 10D coordinates. In (3.7a), we have $\delta_Q A_\pm = 0$ due to the property of $\displaystyle{\not}n$ and $P_\uparrow$.

Our 12D system has a peculiar gauge symmetry which is understood as the $N \to \infty$ reminiscent of the original 10D system. They are dictated with the infinitesimal parameter $\Lambda$ by

$$\delta_G A_\mu = +\Lambda_{,\mu} + g(A_{\mu,+}\Lambda_{,-} - A_{\mu,-}\Lambda_{,+}) ,$$
$$\delta_G \lambda = -g(\Lambda_{,+}\lambda_{,-} - \Lambda_{,-}\lambda_{,+}) . \quad (3.8)$$

Clearly, the terms with $\pm$ are the Poisson brackets with respect to our extra coordinates $\pm$, as the $N \to \infty$ limit of the usual $U(N)$ commutators. Accordingly, $F_{\mu\nu}$ and $D_\mu \lambda$ transform

$$\delta_G F_{\mu\nu} = -g(\Lambda_{,+} F_{\mu\nu,-} - \Lambda_{,-} F_{\mu\nu,+}) ,$$
$$\delta_G (D_\mu \lambda) = -g[\Lambda_{,+}(D_\mu \lambda)_{,-} - \Lambda_{,-}(D_\mu \lambda)_{,+}] . \quad (3.9)$$

These are nothing but the $N \to \infty$ limit of the the $U(N)$ commutators in $\delta_G F_{\mu\nu} = -g[\Lambda, F_{\mu\nu}]$, $\delta_G(D_\mu \lambda) = -g[\Lambda, D_\mu \lambda]$ for a finite $N$. Relevantly, we can confirm the closure of two gauge transformations (3.8) on $A_\mu$:

$$[\delta_G^1, \delta_G^2] A_\mu = g D_\mu(\Lambda^1_{,+}\Lambda^2_{,-} - \Lambda^1_{,-}\Lambda^2_{,+}) , \quad (3.10)$$

where $D_\mu$ contains again the Poisson bracket terms, as in (3.6). From these features, there seems to be no fundamental problem to interpret $F_{\mu\nu}$ and $D_\mu$ as 'field strength' and 'covariant derivatives' in our peculiar 12D space-time.

Our field equations for $A_\mu$ and $\lambda$ are

$$D_j F^{ij} + 2g(\overline\lambda_{,+} P_\uparrow \gamma^i \not{n} \lambda_{,-}) = 0 ~, \tag{3.11}$$

$$\not{n} \gamma^\mu \not{n} D_\mu \lambda = 0 ~. \tag{3.12}$$

Note that the index $\mu$ in (3.12) effectively takes only 10D values, due to the property of $\not{n}$ and $\not{\overline{n}}$.

We can also easily confirm the closure of two supersymmetries in 12D, which is similar to [9]. Our result is

$$[\delta_Q(\epsilon_1), \delta_Q(\epsilon_2)] = \delta_\xi + \delta_\Omega + \delta_\alpha ~, \tag{3.13}$$

where $\delta_\xi$ is for the usual leading translation term, while $\delta_\Omega$ and $\delta_\alpha$ are extra symmetries with respective parameters $\xi^\mu$, $\Omega$ and $\alpha$:

$$\delta_\xi A_\mu = \xi^\nu A_{\mu,\nu} ~, \quad \delta_\xi \lambda = \xi^\nu \lambda_{,\nu} ~, \quad \delta_\Omega A_\mu \equiv \Omega n_\mu ~, \quad \delta_\alpha \lambda \equiv P_\uparrow \alpha ~, \tag{3.14}$$

$$\xi^\mu \equiv (\overline\epsilon_1 P_\uparrow \gamma^{\mu\nu} P_\downarrow \epsilon_2) n_\nu = -(\overline\epsilon_2 P_\uparrow \gamma^{\mu\nu} P_\downarrow \epsilon_1) n_\nu ~,$$

$$\Omega \equiv -\tfrac{1}{2}(\overline\epsilon_1 P_\uparrow \gamma^{ij} P_\downarrow \epsilon_2) F_{ij} = +\tfrac{1}{2}(\overline\epsilon_2 P_\uparrow \gamma^{ij} P_\downarrow \epsilon_1) F_{ij} ~,$$

$$\alpha \equiv \xi^\mu D_\mu \lambda ~. \tag{3.15}$$

Our action (3.5) is of course invariant under these extra symmetries. As usual, the closure on $\lambda$ in (3.13) holds only up to the $\lambda$-field equation (3.12).

Despite of the peculiar property of our gauge symmetry, we can also confirm the Bianchi identity (BI):

$$D_\rho F_{\sigma\tau} + D_\sigma F_{\tau\rho} + D_\tau F_{\rho\sigma} \equiv 0 ~. \tag{3.16}$$

Here the covariant derivative $D_\rho$ contains the Poisson bracket term with $\pm$-derivatives. Relevantly, the following arbitrary variations for $F_{\mu\nu}$ and $D_\mu \lambda$ are useful:

$$\delta F_{\mu\nu} = D_\mu(\delta A_\nu) - D_\nu(\delta A_\mu) ~,$$

$$\delta(D_\mu \lambda) = D_\mu(\delta \lambda) + g[\,(\delta A_\mu)_{,+} \lambda_{,-} - (\delta A_\mu)_{,-} \lambda_{,+}\,] ~. \tag{3.17}$$

Needless to say, all of these covariant derivatives contain the $\pm$-derivatives. Note also that $D_\mu$ satisfies the Leibnitz rule: $D_\mu(AB) = (D_\mu A)B + A(D_\mu B)$, enabling us to perform partial integrations under $\int d^{12}x$. Using this with (3.17), it is now straightforward to obtain our field equations (3.11) and (3.12), and also to confirm the invariance of our total action $\delta_Q I = 0$ under our supersymmetry (3.7). The transformation rule $\delta_Q A_\pm = 0$ in (3.7) poses no problem, due to the effective absence of $A_\pm$ in our action (3.5).

Even though the extra components $F_{\pm i}$ and $F_{+-}$ are effectively absent from our action (3.5), the system is *not* reduced to just an infinite identical copies of supersymmetric YM theory in 10D, or a rewriting of the latter 'in disguise'. This is due to the non-trivial Poisson bracket terms in $F_{ij}$ which are the non-trivial reminiscent of the non-Abelian terms in the original 10D theory.

One crucial question is whether or not our 12D theory can be Lorentz covariant. Even though we still lack a Lorentz invariant lagrangian yet, we emphasize as in [16][13] that all of our field equations can be made entirely Lorentz covariant, by expressing the null-vectors in terms of two scalars: $n_\mu \equiv \varphi_{,\mu}$, $m_\mu \equiv \widetilde{\varphi}_{,\mu}$, satisfying $\varphi_{,\mu\nu} = 0$, $\widetilde{\varphi}_{,\mu\nu} = 0$, $(\varphi_{,\mu})^2 = 0$, $(\widetilde{\varphi}_{,\mu})^2 = 0$, $\varphi_{,\mu} \widetilde{\varphi}_{,}{}^\mu = +1$ [16][13]. From this viewpoint, our system has another non-trivial feature, even for Lorentz covariance. The possibility of a Lorentz invariant lagrangian is now under study.

### 4. Superspace Formulation in 12D

Once we have understood the component formulation of our 12D system, the next natural task is to reformulate this system in superspace. This is done mainly by studying the geometrical significance of supercovariant derivatives, and satisfaction of BIs for superfield strength in superspace.

Our superspace coordinates are $(Z^A) = (x^a, \theta^\alpha)$, where we use the superspace index convention: $A = (a,\alpha)$, $B = (b,\beta)$, $\cdots$, with the bosonic coordinates $a, b, \cdots = 0, 1, \cdots, 9, +, -$ and the fermionic coordinates $\alpha, \beta, \cdots = 1, 2, \cdots, 32$. As usual, our starting point is the super-gauge covariant derivative, defined in our case by

$$\nabla_A \equiv D_A + g(A_{A,+} D_- - A_{A,-} D_+) \ , \tag{4.1}$$

where $D_A \equiv E_A{}^M \partial_M$ is the usual superspace covariant derivative, while $A_{A,\pm} \equiv D_\pm A_A \equiv \partial_\pm A_A$. We regard the last two terms in (4.1) as a 'gauge connection' term, generating the Poisson bracket terms in the superfield strength, as will be seen. Note that only $D_\pm$ instead of $\nabla_\pm$ are needed for these terms. Now our superfield strength $F_{AB}$ is defined by the commutator as

$$[\nabla_A, \nabla_B\} = g(F_{AB,+} D_- - F_{AB,-} D_+) \ , \tag{4.2a}$$

$$F_{AB} \equiv D_{[A} A_{B)} - T_{AB}{}^C A_C + g(A_{A,+} A_{B,-} - A_{A,-} A_{B,+}) \ , \tag{4.2b}$$

where the last Poisson bracket terms are similar to the component case (2.6), as the reminiscent of the commutator of $U(N)$ generators in the large $N$ limit. From the commutator defining $F_{AB}$, we see the first signal of the geometric significance of our formulation. Even though it may be expected in a certain sense, it is remarkable that our new superfield strength $F_{AB}$ satisfies the BIs

$$\nabla_{[A} F_{BC)} - T_{[AB|}{}^D F_{D|C)} \equiv 0 \ , \tag{4.3}$$

$$\nabla_{[A} T_{BC)}{}^D - T_{[AB|}{}^E T_{E|C)}{}^D \equiv 0 \ , \tag{4.4}$$

where (4.4) is the BI for the torsion superfield $T_{AB}{}^C$, and $\nabla_A$ acts on $F_{BC}$ with the Poisson bracket terms as in (4.1). These BIs are confirmed by the Jacobi identity $[[\nabla_{[A}, \nabla_B\}, \nabla_{C)}\} \equiv 0$. To specify the components in (4.3), we call it $(ABC)$-type BI.

Our next task is to satisfy all the components of the BI (4.3) and (4.4). As usual in superspace formulations, we postulate a set of constraints:

$$T_{\alpha\beta}{}^c = (\gamma^{cd})_{\alpha\beta} n_d + (P_{\uparrow\downarrow})_{\alpha\beta} n^c = (P_\uparrow \gamma^{cd} P_\downarrow)_{\alpha\beta} n_c \ , \tag{4.5a}$$

$$F_{ab} = -(P_\uparrow \gamma_b \slashed{n} \lambda)_\alpha + n_b \chi_\alpha \ , \tag{4.5b}$$

$$\nabla_\alpha \lambda_\beta = -\tfrac{1}{4}(P_\uparrow \gamma^c P_\uparrow \gamma^d P_\uparrow)_{\alpha\beta} F_{cd} \ , \tag{4.5c}$$

$$\nabla_\alpha \chi_\beta = +\tfrac{1}{2}(P_\uparrow \gamma^a \slashed{n})_{\alpha\beta} F_{ab} m^b \ , \tag{4.5d}$$

$$\nabla_\alpha F_{bc} = +(P_\uparrow \gamma_{[b|} \slashed{n} \nabla_{|c]} \lambda)_\alpha - n_{[b} \nabla_{c]} \chi_\alpha \ . \tag{4.5e}$$

There is similarity as well as difference between this superspace formulation and that in [9][17] or that in [14]. First, (4.5a) is exactly the same as that in [14], in particular with the $P_{\uparrow\downarrow}$-term. Another similarly is that the $\chi$-field is a kind of auxiliary field, needed for the $(\alpha\beta c)$-type BI, but it disappears from the final superfield equations (4.7) and (4.8) below. The difference is that the fields $\lambda$ or $F_{AB}$ are *not* subject to any extra constraints, such as $F_{ab} n^b = 0$ [9][17].

The satisfaction of BI (4.4) is rather trivial, due to the non-vanishing component of $T_{\alpha\beta}{}^c$ in (4.5a). The confirmation of the BI (4.3) is as easy as the other 12D cases [9][17], up to some points peculiar to this system. First, the $(\alpha\beta\gamma)$-type BI is proportional to

$$(\gamma^i \slashed{n})_{(\alpha\beta|} (\gamma_i \slashed{n} \lambda)_{|\gamma)} \equiv 0 \ , \tag{4.6}$$

confirmed by 12D $\gamma$-matrix algebra as in [9][13]. The $(\alpha\beta c)$-type BI is easily shown to be satisfied due to (4.5c), while $(\alpha b c)$-type BI gives (4.5e). At this stage, we can get the $\lambda$-superfield equation as $\nabla_{(\alpha}(\nabla_{\beta)}\lambda^\beta) - \{\nabla_\alpha, \nabla_\beta\}\lambda^\beta \equiv 0$. This $\lambda$-superfield equation in turn gives the $F$-superfield equation by taking its spinorial derivative like $(\gamma^a)^{\alpha\gamma}\nabla_\gamma(\lambda_\alpha\text{-field equation}) = 0$, both in agreement with (3.11) and (3.12). In the present notation they are

$$\nabla_j F^{ij} + 2g\lambda^\alpha{}_{,+}(\gamma^i \not{\eta})_\alpha{}^\beta \lambda_{\beta,-} = 0 \ , \tag{4.7}$$

$$(\not{\eta}\gamma^a\not{\eta})_\alpha{}^\beta \nabla_a \lambda_\beta = 0 \ . \tag{4.8}$$

As has been already mentioned, even though the extra components $F_{\pm i}$ and $F_{+-}$ are absent in (4.7), the Poisson bracket terms with extra derivatives $_{,\pm}$ in these superfield equations differentiate our 12D system from merely a rewriting of 10 supersymmetric theory, or the latter just in disguise.

It is interesting that our newly-defined superfield strength $F_{AB}$ reveals so much geometrical significance and consistency with supersymmetry, quite parallel to conventional superspace formulations. This already suggests much deeper physical and geometrical significance of the incorporation of the symplectic canonical variables $p$, $q$ as a part of the space-time coordinates, forming the total space-time with two time coordinates.

## 5. Dimensional Reduction into 3D

As an important test of our 12D theory, we perform a dimensional reduction into 3D, and see if the resulting action coincides with the $N \to \infty$ limit of $D0$-brane action in 1D [1]. This is because the the two canonical symplectic variables $p$, $q$ parametrizing the unitary $N \times N$ matrices in 1D form additional two extra coordinates, promoting it to a 3D theory, compatible with the action (hamiltonian) [11] of supermembrane [18].

Our dimensional reduction prescription is the usual one, namely we require all the fields to be independent of the internal 9D coordinates $x^1, x^2, \cdots, x^9$, so that the space-integrals over these coordinates in the action become an over-all trivial factor. Only in this section we use *hats* on fields and on indices in 12D in order to distinguish them from those in 3D. In this notation, our 12D lagrangian (3.5) is rewritten as

$$I = \int d^{12}\widehat{x} \left[ -\tfrac{1}{4}(\widehat{F}_{\widehat{i}\widehat{j}})^2 + (\overline{\widehat{\lambda}}\widehat{P}_\uparrow \widehat{\gamma}^{\widehat{i}}\not{\eta}\widehat{D}_{\widehat{i}}\widehat{\lambda}) \right] \ , \tag{5.1}$$

where $\widehat{i}, \widehat{j}, \cdots = 0, 1, \cdots, 9$ are 10D coordinate indices. Other components in $\widehat{F}_{\widehat{\mu}\widehat{\nu}}$ such as $\widehat{F}_{\widehat{i}+}$ are effectively absent from (5.1), due to the second and third terms in (3.5). All the 12D fields are dimensionally reduced to 3D with the coordinates $(x^0, x^+, x^-)$ by the rules, such as

$$\widehat{F}_{\widehat{i}\widehat{j}} = \begin{cases} \widehat{F}_{ij} = g(X^i{}_{,+}X^j{}_{,-} - X^i{}_{,-}X^j{}_{,+}) \ , \\ \widehat{F}_{0i} = X^i{}_{,0} + g(A_{0,+}X^i{}_{,-} - A_{0,-}X^i{}_{,+}) \equiv D_0 X^i \ , \end{cases} \tag{5.2a}$$

$$\widehat{\lambda} = \begin{pmatrix} 0 \\ \lambda \end{pmatrix} \ , \quad \overline{\widehat{\lambda}} = (0, \overline{\lambda})\, (I_{32} \otimes \tau^1) = (\overline{\lambda}, 0) \ , \tag{5.2b}$$

$$\widehat{\gamma}^{\hat{i}} = \begin{cases} \widehat{\gamma}^i = \Gamma^i \otimes \tau^3 \ , \\ \widehat{\gamma}^0 = \Gamma^0 \otimes \tau^3 \ , \\ \widehat{\gamma}^+ = I_{32} \otimes \tau^+ \ , \\ \widehat{\gamma}^- = I_{32} \otimes \tau^- \ , \end{cases} \quad \widehat{P_\uparrow} = \begin{pmatrix} I_{32} & 0 \\ 0 & 0 \end{pmatrix} \ , \quad \widehat{P_\downarrow} = \begin{pmatrix} 0 & 0 \\ 0 & I_{32} \end{pmatrix} \ , \tag{5.2c}$$

$$\widehat{\eta} = I_{32} \otimes \tau^+ \ , \quad \widehat{\overline{\eta}} = I_{32} \otimes \tau^- \ , \tag{5.2d}$$

where $i, j, \cdots = 1, \cdots, 9$ in this section are for the spacial 9D, as those used in [1]. Therefore $(\Gamma^0, \Gamma^i)$ in (5.2c) realize the Clifford algebra for 10D. In (5.2), $\lambda$ is a 10D Majorana-Weyl spinor with maximally 32 components, and $\tau^1$, $\tau^2$, $\tau^3$ are the usual $2 \times 2$ Pauli matrices, and $\tau^\pm \equiv (\tau^1 \pm i\tau^2)/2$. Applying the useful relations (5.2) to the action (5.1), and integrating over the internal 9D coordinates $x^1, \cdots, x^9$, we get

$$I = \int d^3x \left[ + \tfrac{1}{2}(D_0 X^i)^2 - \tfrac{1}{4}g^2 (X^i{}_{,+} X^j{}_{,-} - X^i{}_{,-} X^j{}_{,+})^2 \right.$$
$$\left. + \sqrt{2}\, g\, \overline{\lambda} \Gamma^i (X^i{}_{,+} \lambda_{,-} - X^i{}_{,-} \lambda_{,+}) + \sqrt{2}\, (\overline{\lambda} \Gamma^0 D_0 \lambda) \right] \ . \tag{5.3}$$

This lagrangian is still in terms of 10D spinor $\lambda$, which is to be further reduced to 16 component spinors to fit the $SO(9)$ symmetry we need to compare with the result in [1]. This can be done by the $SO(9)$ $\gamma$-matrix representations [1][5]:

$$\lambda = \begin{pmatrix} \theta \\ 0 \end{pmatrix} \ , \quad \overline{\lambda} = (0, \theta^{\mathrm{T}}) \ , \quad \Gamma^0 = \begin{pmatrix} 0 & -I_{16} \\ I_{16} & 0 \end{pmatrix} \ , \quad \Gamma^i = \begin{pmatrix} 0 & \gamma^i \\ \gamma^i & 0 \end{pmatrix} \ , \tag{5.4}$$

yielding the lagrangian

$$I = \int d^3x \left[ + \tfrac{1}{2}(D_0 X^i)^2 - \tfrac{1}{4}g^2 (X^i{}_{,+} X^j{}_{,-} - X^i{}_{,-} X^j{}_{,+})^2 \right.$$
$$\left. + \sqrt{2}\, g\, \theta^{\mathrm{T}} \gamma^i (X^i{}_{,+} \theta_{,-} - X^i{}_{,-} \theta_{,+}) + \sqrt{2}\, (\theta^{\mathrm{T}} D_0 \theta) \right] \ . \tag{5.5}$$

Accordingly, we can perform the similar dimensional reduction to our supersymmetry transformation (3.7), to get

$$\delta_Q X^i = -\sqrt{2}\, (\epsilon^{\mathrm{T}} \gamma^i \theta) \ , \quad \delta_Q A_0 = \sqrt{2}\, (\epsilon^{\mathrm{T}} \theta) \ ,$$
$$\delta_Q A_+ = \delta_Q A_- = 0 \ ,$$
$$\delta_Q \theta = -\tfrac{1}{2} \gamma^i \epsilon D_0 X^i + \tfrac{1}{4} g \gamma^{ij} \epsilon\, (X^i{}_{,+} X^j{}_{,-} - X^i{}_{,-} X^j{}_{,+}) \ . \tag{5.6}$$

Our results (5.5) and (5.6) agree with the 3D result in [11] after the large $N$ limit, up to non-essential scaling factors such as $\sqrt{2}$.

We have thus seen that our 12D supersymmetric YM theory directly gives rise to the 3D theory [11] corresponding to the supermembrane theory [18] after taking the large $N$ limit in the $D0$-brane action.

## 6. Concluding Remarks

In this paper we have presented a 12D supersymmetric gauge theory, with a very peculiar gauge symmetry associated with Poisson bracket as a reminiscent of the commutators in non-Abelian generators in the original 10D theory. The Poisson bracket arises in the 't Hooft $N \to \infty$ limit for $U(N)$, whose $p$, $q$ variables are now regarded as two extra coordinates in the total 12D. The extra coordinates have the ranges $-\pi \leq p < \pi$, $-\pi \leq q < \pi$, implying a compactification on $S^1 \otimes S^1$ within the total 12D. These extra dimensions are further decompactified by the strong/weak duality by rescaling the coordinates by $x^{11} \equiv Rp$, $x^{12} \equiv Rq$, and taking the limit $R \to \infty$, $g \to 0$ with $R^2 g$ fixed. Subsequently, we have also studied the geometrical significance in superspace of our superfield strength with the Poisson brackets, and found that a superspace formulation is equally possible like the conventional supersymmetric YM theory. This superspace formulation elucidates the geometrical significance of our theory, in terms of field strength superfield defined by commutators between super-gauge covariant derivatives, with our peculiar 'gauge' symmetry. We have also studied the dimensional reduction of our 12D theory into 3D, that yields the desirable Poisson brackets in 3D, corresponding to the action for supermembrane theory [18][11].

In a recent development in the duality between Anti-de Sitter (AdS) and conformal field theory [19], it is conjectured that the $\mathcal{N} = 4$, $U(N)$ supersymmetric YM theory in 4D is dual to type IIB theory on $(\text{AdS})_5 \otimes S^5$ in the large $N$ limit. It was pointed out [19] that the group $SO(2,4) \times SO(6)$ in the $\mathcal{N} = 4$ supersymmetric YM theory suggests a 12D realization with two time coordinates. In our present paper, we have given a first explicit example, in which the direct connection between the supermembrane theory and 10D supersymmetric YM is much more natural than before. Since the $\mathcal{N} = 4$ supersymmetric YM in 4D has the 10D origin which is promoted to be 12D in the large $N$ limit, we have seen another duality link between superstring theory [3], $\mathcal{N} = 4$ supersymmetric YM in 4D, and supermembrane theory [18] *via* 12D supersymmetric YM theory [9].

We have observed similarity as well as difference between our new superspace formulation and that in [9][17]. The most conspicuous difference is the non-vanishing Poisson bracket terms in $F_{AB}$ making the system non-trivial. We have also clarified the geometrical significance of our peculiar field strength superfield in terms of superspace language. We also stress that our present paper gives a clearer link between supersymmetric YM theory in 12D [9][17] and M-theory [4][5].

In a usual non-supersymmetric theory, the loss of Lorentz covariance makes the system completely ambiguous, because we can always put null-vectors anywhere by hand in any equation in the system. However, as was also stressed in [13][17], this is no longer the case with supersymmetric theories, in which the coefficients in field equations are tightly fixed by supercovariance. The incorporation of symplectic variables as target coordinates makes stronger sense in supersymmetric theories. Our formulation has been also strongly motivated by the recent development in the matrix theory approach [1] to M-theory [4][5].

In our formulation, due to the absence of supergravity, the loss of local Lorentz covariance is not crucial. In this connection, we mention that only global Lorentz covariance plays an important role in the study of non-perturbative aspects of M-theory. In fact, in the conjecture by Maldacena [19] about the duality between the large $N$ limit of $D=4$, $\mathcal{N}=4$ supersymmetric Yang-Mills and type IIB superstring compactified on $(AdS)_5 \otimes S^5$, the *global* isometry group is $SO(4,2) \times SO(6)$, being further promoted to $SO(10,2)$ [19][20], indicates the existence of 12D supersymmetric theory. Note that the $SO(10,2)$ symmetry is global, and therefore the loss of *local* Lorentz covariance in 12D supergravity/supersymmetry formulations as in [13] is not crucial in such a formulation of M-theory [4][5].

Our result also provides a new link between the F-theory [21] in 12D and the M-theory [4] in 11D, that has not been explicitly presented before.[7] Our result also clarifies general dualities connecting a supersymmetric YM theory in $(D-1,1)$ dimensions with another supersymmetric theory formulated in $(D,2)$ dimensions under the large $N$ limit. It also gives a new connection between the conventional supersymmetry/supergravity theories in $D \leq 11$ with higher-dimensional theories, such as the 12D supergravity [14][15][13], or more supersymmetric YM theories in $D \geq 12$ [17]. In particular, the evidence for the importance of the results in [9][17] is now rapidly mounting.

We are indebted to I. Bars, S.J. Gates, Jr., J.H. Schwarz, and W. Siegel for valuable suggestions.

---

[7]In ref. [22], an idea of deleting even the world-line for the $D0-$brane was presented, but with no reference to the 12D YM field strength defined with Poisson bracket in terms of extra coordinates.